\documentclass[aps,twocolumn,nofootinbib,showpacs, prd]{revtex4}
\usepackage{graphicx}
\usepackage{amsfonts}
\usepackage{amssymb}
\usepackage{amsbsy}
\usepackage{amsmath}
\usepackage{latexsym}
\usepackage{natbib}
\usepackage{bm}
\usepackage{color}
\usepackage{float}

\begin{document}
\title{\bf Collapsing spherical star in Scalar-Einstein-Gauss-Bonnet gravity with a quadratic coupling}

\author{Soumya Chakrabarti}
\email{soumya@cts.iitkgp.ernet.in}

\affiliation{Centre for Theoretical Studies,\\
Indian Institute of Technology Kharagpur, \\
Kharagpur 721 302, West Bengal, India
} 

\date{\today}

\begin{abstract}
We study the evolution of a self interacting scalar field in Einstein-Gauss-Bonnet theory in four dimension where the scalar field couples non minimally with the Gauss-Bonnet term. Considering a polynomial coupling of the scalar field with the Gauss-Bonnet term, a self-interaction potential and an additional perfect fluid distribution alongwith the scalar field, we investigate different possibilities regarding the outcome of the collapsing scalar field. The strength of the coupling and choice of the self-interaction potential serves as the pivotal initial conditions of the models presented. The high degree of non-linearity in the  equation system is taken care off by using a method of invertibe point transformation of anharmonic oscillator equation, which has proven itself very useful in recent past while investigating dynamics of minimally coupled scalar fields.
\end{abstract}
\maketitle

\section{Introduction}
General theory of relativity is widely regarded to be the best theory of gravity as over the years it has passed many observational tests with flying colors. Significant observational aspects such as the perihelion precission of Mercury's orbit, gravitational lensing, redshift in the light spectrum from extragalactic objects are well documented in the framework of general relativity. However, general relativity can pose considerable intrigue in certain aspects such as the possibility of an extremely strong gravitational field imploding without limit and ending up in a region where the density of matter and the strength of the gravitational field can in principle, become infinite. Such a region is called a spacetime singularity. Aspects of gravitational collapse and the formation of a spacetime singularity forms an integral part of gravitational physics today. \\

In general, a continual gravitational collapse occurs whenever a massive astronomical body, upon exhausting it's nucleur fuel supply, fails to support itself against the force of gravity. For a simple enough configuration of collapsing matter, a horizon generally develops prior to the formation of the singularity, thereby enveloping the singularity producing a black hole end-state. The first analytic model of such an unhindered contraction of an idealized star ending up in a black hole was given by Oppenheimer and Snyder \cite{os} and independently by Dutt \cite{dutt} and these serve as the paradigm of gravitational collapse today. However, whether or not every sufficiently massive star undergoing gravitational collapse ultimately ends up in a black hole, remains an intriguing question even in the current context. In this regard, Penrose proposed \cite{rp} the cosmic censorship hypothesis (CCH), which roughly states that gravitational collapse of physically reasonable matters with generic regular initial data will always end up in a covered singularity. 
\\

The CCH, however, remains one of the most thought-provoking problems in gravitational physics today. There is no general proof of the CCH as yet which applies under all conditions. Moreover, there are many counterexamples in general relativity, where it is shown that the singularities formed in a collapse of reasonable distribution of matter can stay exposed, giving rise to the concept of a naked, i.e., an observable singularity (for a brief summary of such examples, we refer to \cite{thesis} and references therein). This much is realized that the nature of the final outcome of a collapse depends on the initial configuration from which the collapse evolves, and the allowed dynamical evolutions in the spacetime, as permitted by the non-linear field equations of gravity. For recent works and different aspects regarding the dynamics of a continued gravitational collapse we refer to the summary by Joshi \cite{joshi1, joshi2}. \\

The singularity or a spacetime region with infinitely high curvature can be realized only during the final stage of gravitational collapse, where the collapsing body almost reaches a zero proper volume and known laws of physics are expected to break down. A possibility of naked singularity implies that one may have a chance to observe indication of quantum effects of gravity. Amongst the many quantum theories of gravity proposed, superstring/M-theory is a promising candidate, which motivates the presence of a higher-dimensional spacetime \cite{M}. During the final stages of the stellar evolution, where the curvature of the central high-density region is very high, the effects of extra dimensions can perhaps play a crucial role. From such a perspective, higher-dimensional gravitational collapse models are studied in general relativity (we refer to the works of Banerjee, Debnath and Chakraborty \cite{bdc}, Patil \cite{patil}, Goswami and Joshi \cite{gosjos} in this regard). Given that the entire aspects of superstring/M-theory are not understood completely so far, taking their effects perturbatively into classical gravity is one possible approach to study higher curvature effects. The Gauss-Bonnet (GB) term, $G = R^2 - 4R_{\mu\nu}R^{\mu\nu} + R_{\mu\nu\alpha\beta}R^{\mu\nu\alpha\beta}$ in the standard Einstein-Hilbert Lagrangian is the higher curvature correction to general relativity which finds it's motivation from heterotic superstring theory \cite{superstring}. Such a theory is called the Einstein-Gauss-Bonnet gravity.  \\

The additional elegance of including the GB term is that this is a Lovelock scalar and if included linearly in the action, the field equations include no higher than second order partial derivatives of the metric tensor (unlike $f(R)$ gravity whose field equations are fourth order in metric components). In a four dimensional spacetime the Gauss-Bonnet term does not modify the field equations. However, if the non-minimal coupling of a scalar field with the GB term is considered, the dynamical equations are quite different from the standard field equations and the influence of the GB term in a four dimensional universe is effective. \\

Recently there is an increasing interest in Gauss-Bonnet theory with a non-minimally coupled scalar field to suffice for the possible candidature of the late time acceleration of the universe \cite{nojiri1, nojiri2, cognola, koiv1, koiv2}. From such a perspective, spherically symmetric solutions has been studied by Boulware and Deser\cite{boul1, boul2} and Gurses \cite{gurses}. It has also been discussed that the effective action including correction terms of higher order in the curvature can perhaps play a significant role in the dynamics of early universe by Zwiebach\cite{zwi}, Zumino\cite{zu}, Boulware and Deser\cite{boul1, boul2}. Questions regarding gravitational instability, cosmological perturbation were also considered by Kawai, Sakagami and Soda\cite{kawa1, soda}, Kawai and Soda\cite{kawa2}. Observational restrictions over different cosmological aspects of the scalar field coupled Einstein Gauss Bonnet gravity were investigated by Guo and Schwarz\cite{guo}, Koh et. al.\cite{koh}. Spherically symmetric collapsing solutions of this theory have also gained some interest quite recently. For instance, Maeda presented a model of $n \geq 5$ dimensional spherically symmetric gravitational collapse of a null dust fluid in Einstein-Gauss-Bonnet gravity \cite{maeda1} and illustrated the possibility of a formation of massive naked singularity in higher dimensions. He comparitively analyzed the results with the general relativistic cases as well \cite{maeda2}, which serves as a higher order generalization of the Misner-Sharp formalism of the four-dimensional spherically symmetric spacetime with a perfect fluid in general relativity. Hamiltonian Formulation of spherically symmetric scalar field collapse in Gauss-Bonnet gravity was studied in detail by Taves, Leonard, Kunstatter and  Mann \cite{mann1}. They also proved that such a formulation can readily be generalized to other matter fields minimally coupled to gravity. Apart from their role in cosmology, the role of scalar fields in a gravitational collapse is worthy of attention. It is indeed important to investigate if the CCH necessarily holds or violates in the collapse of fundamental matter fields. Moreover, a scalar field alongwith an interaction potential is known to mimic different kind of reasonable distribution of matter as discussed by Goncalves and Moss \cite{gonca}. Numerical simulations of the spacetime evolution of a massless scalar field minimally coupled to gravitational field studied by Choptuik \cite{chop}, Brady \cite{brady} and Gundlach \cite{gund} hint interesting possibilities, such as, the critical behavior observed around the threshold of black hole formation. There is a self-similar solution that sits at the black hole threshold, dubbed a critical solution, and also a very interesting mass scaling law for the formed black hole end-state. Motivated by this, self similar collapsing scenario of a massive scalar field was analytically studied by Banerjee and Chakrabarti \cite{scnb1} very recently. Deppe, Taves, Leonard, Kunstatter and  Mann presented a numerical analysis in generalized flat slice co-ordinates of self-gravitating massless scalar field collapse in five and six dimensional Einstein Gauss Bonnet gravity near the threshold of black hole formation \cite{mann2}. Effects of higher order curvature corrections to Einstein's Gravity on the critical phenomenon near the black hole threshold, were investigated by Golod and Piran \cite{golod}. In general relativity also, possibilities and scope of scalar field collapse have been analytically studied extensively \cite{scalarcollapse}. Finally, the fact that the distribution of the dark energy component or the fluid still remains unknown, naturally inspires a continuing study of scalar field collapse under increasingly generalized setup (preferably alongwith a fluid distribution) towards a better understanding of the possible clustering of dark energy. \\

In this work, we study aspects of a scalar field collapse in a Scalar-Einstein-Gauss-Bonnet gravity, where the self-interacting scalar field $\phi$ is non-minimally coupled to the GB term. Very recently Banerjee and Paul \cite{nbtp} studied such a scalar field collapse where the coupling term was proportional to $e^{2\phi}$. In the present case the coupling term is proportional to $\phi^2$, i.e., the coupling is quadratic in $\phi$. Quite recently, Doneva and Yazadjiev showed that in a very similar setup of Scalar Einstein Gauss Bonnet theory (the conditons they imposed on the coupling function $f(\phi)$ are $f'(\phi = 0) = 0$ and $b^2 = f''(\phi = 0) > 0$), there exists new black hole solutions which are formed by spontaneous scalarization of the Schwarzaschild black holes in the extreme curvature regime and below a certain mass, the Schwarzschild solution becomes unstable and new branch of solutions with nontrivial scalar field bifurcate from the Schwarzschild one \cite{doneva1}. They also proved the existence of neutron stars in a class of extended scalar-tensor Gauss-Bonnet theories for which the neutron star solutions are formed via spontaneous scalarization of the general relativistic neutron stars \cite{doneva2}. Very recently, the spontaneous scalarization of black holes and compact stars from such a Gauss-Bonnet coupling has been investigated and dubbed as the Quadratic Scalar-Gauss-Bonnet gravity by Silva et. al. \cite{silva}. In this context, existence of regular black-hole solutions with scalar hair in the Einstein-scalar-Gauss-Bonnet theory was investigated by Antoniou, Bakopoulos and Kanti \cite{kanti0, kanti00}, with a general coupling function between the scalar field and the quadratic Gauss-Bonnet term which highlighted the limitations of the existing no-hair theorems. In a recent study Kanti, Gannouji and Dadhich have addressed the importance of such a coupling from a cosmological purview and proved by some simple analytical calculation that a quadratic coupling function, although a special choice, allows for inflationary, de Sitter-type solutions to emerge \cite{kanti}.  \\

The inclusion of the Gauss-Bonnet terms make the dynamical field equations even more non-linear. We study a spatially homogeneous model where the energy momentum tensor is contributed by the self-interacting scalar field as well as a perfect fluid. To track down the system of equatons, we use the method of invertible point transformations and integrability of anharmonic oscillator equations; an approach which has been quite useful recently, in the study of minimally coupled massive scalar field collapse by Chakrabarti and Banerjee \cite{scnb1, scnb2}.  \\

The paper is organized as follows. In section $II$, we introduce the action and basic field equations. Section $III$ contains the method of finding the exact solution and in sections $IV$ we study the evolution of the scale factor for different initial data. The time evolution of the scalar field is studied in section $V$. Evolution of the curvature scalars and the strong energy condition is studied in section $VI$. The physical nature of the singularity is addressed in $VII$. We complete the model by matching the solution with an exterior Vaidya solution in section $VIII$ and conclude in section $IX$.  

\section{Action and Basic Equations}
The relevant action for a four dimensional action containing the Einstein-Hilbert part, massive scalar field and the Gauss-Bonnet term coupled with the scalar field. The corresponding action is given by  
\begin{eqnarray}\nonumber\nonumber
\label{action}
&& S = \int d^4x \sqrt{-g} \bigg[R/(2\kappa^2)-(1/2)g^{\mu\nu}\partial_{\mu}\phi\partial_{\nu}\phi-V(\phi)\\&&\nonumber
 - \xi(\phi)G \bigg],
 \label{action}\\
\end{eqnarray}
where $R$ is the Ricci scalar, $1/(2\kappa^2) = M_{p}^2$ is the four dimensional squared Planck scale and $G = R^2 - 4R_{\mu\nu}R^{\mu\nu} + R_{\mu\nu\alpha\beta}R^{\mu\nu\alpha\beta}$ is the GB term. $\phi$ and $V(\phi)$ denote the scalar field and the self interaction potential respectively. $\xi(\phi)$ defines the coupling between scalar field and GB term. Variation of the action with respect to metric and scalar field leads to the field equations as follows,
\begin{eqnarray}
&\frac{1}{\kappa^2}&[-R^{\mu\nu}+(1/2)g^{\mu\nu}R] + (1/2)\partial^{\mu}\phi\partial^{\nu}\phi \nonumber\\ 
&-& (1/4)g^{\mu\nu}\partial_{\rho}\phi\partial^{\rho}\phi + (1/2)g^{\mu\nu}\big[-V(\phi)+\xi(\phi)G\big] \nonumber\\ 
&-& 2\xi(\phi)RR^{\mu\nu} - 4\xi(\phi)R^{\mu}_{\rho}R^{\nu\rho} - 2\xi(\phi)R^{\mu\rho\sigma\tau}R^{\nu}_{\rho\sigma\tau} \nonumber\\ 
 &+& 4\xi(\phi)R^{\mu\rho\nu\sigma}R_{\rho\sigma} + 2[\nabla^{\mu}\nabla^{\nu}\xi(\phi)]R-2g^{\mu\nu}[\nabla^2\xi(\phi)]R \nonumber\\ 
&-& 4[\nabla_{\rho}\nabla^{\mu}\xi(\phi)]R^{\nu\rho} - 4[\nabla_{\rho}\nabla^{\nu}\xi(\phi)]R^{\mu\rho} + 4[\nabla^2\xi(\phi)]R^{\mu\nu}\nonumber\\ 
 &+&4g^{\mu\nu}[\nabla_{\rho}\nabla_{\sigma}\xi(\phi)]R^{\rho\sigma} + 4[\nabla_{\rho}\nabla_{\sigma}\xi(\phi)]R^{\mu\rho\nu\sigma} = 0
 \label{mainfieldequation}
 \end{eqnarray}
 and
 \begin{equation}
  g^{\mu\nu}[\nabla_{\mu}\nabla_{\nu}\phi] - V'(\phi) - \xi'(\phi)G = 0,
  \label{scalarfieldequation}
\end{equation}
where a prime denotes the derivative with respect to $\phi$. \\

The metric for the interior is assumed to be a spatially flat Friedmann-Robertson-Walker metric and is given by
\begin{equation}
 ds^2 = -dt^2 + a^2(t)\big[dr^2 + r^2d\theta^2 + r^2\sin^2{\theta}d\varphi^2\big]
 \label{metric}
\end{equation}
where the scale factor $a(t)$ governs the time evolution of the interior spacetime. For such a metric, the expression of Ricci scalar $R$ and GB term $G$ take the following form,
\begin{eqnarray}\label{scalar}
 R = 6[2H^2 + \dot{H}]\nonumber\\
 G = 24H^2[H^2 + \dot{H}]
 \nonumber
\end{eqnarray}
where $H = \dot{a}/a$ and dot denotes the derivative with respect to time $(t)$.  \\

Here, we have assumed the interior of the collapsing star to be consisting of the scalar field as well as a perfect fluid. Therefore, using the metric in eqn. (\ref{metric}), the field equations can be written as,
\begin{equation}
 -(3/\kappa^2)H^2 + (1/2)\dot{\phi}^2 + V(\phi) + 24H^3\dot{\xi} + \rho = 0,
 \label{G00}
\end{equation}
\begin{eqnarray}\nonumber
&& \frac{1}{\kappa^2}\big[2\dot{H}+3H^2\big] + (1/2)\dot{\phi}^2 - V(\phi) - 8H^2\ddot{\xi} \\&&
 - 16H\dot{H}\dot{\xi} - 16H^3\dot{\xi} + p = 0,
 \label{G11}
\end{eqnarray}
and 
\begin{equation}
 \ddot{\phi} +3H\dot{\phi} + V'(\phi) + 24\xi'(\phi)(H^4+H^2\dot{H}) = 0.
 \label{scalarKG}
\end{equation}

$\rho$ and $p$ signifies the density and pressure of the constituent fluid inside the collapsing star.

\section{Exact Solution}
It can be easily noted that, due to the inclusion of the additional fluid component (two additional unknown functions $\rho$ and $p$), the system of equations becomes even more difficult to deal with analytically, even more so without assuming any particular equation of state. However, we try to do away with the difficulty by incorporating a strategy, where the scalar field evolution equation (\ref{scalarKG}) is identified as an anharmonic oscillator equation and integrated straightaway without any apriori assumptions regarding the equation of state or the scale factor. In this way, the other field equations can be used to study the evolution of the fluid/scalar field distribution in a general manner. However, the only assumption being implemented here is that the equation (\ref{scalarKG}) is integrable. The criterion for such an integrability can be defined in terms of an invertible point transformation, first worked out by Duarte et. al. \cite{duarte}, Euler, Steeb and Cyrus \cite{euler}. Although the main motivation over this assumption of integrability is extracting information from the field equations at any cost without resorting to any assumption of equation of state, this assumption is by no means unphysical. In recent past, this approach has proved to be very useful to produce interesting solutions depicting dynamics of minimally coupled massive scalar field \cite{scnb2}, self-similar solutions \cite{scnb1} and also showed promise towards being a tool for reconstructing modified gravity lagrangians. Using this approach, Chakrabarti, Said and Farrugia have, quite recently studied a reconstruction method for teleparrallel gravity \cite{scjs}. This work therefore, carries a subtle motivation of testing the scope of this approach in the domain of Scalar-Einstein-Gauss-Bonnet gravity.    \\

While a simple linear harmonic oscillator has a straightforward sinusoidal solution, an anharmonic oscillator has more contributing terms and can represent more physical features of a dynamical system. This takes the form of a nonlinear second order differential equation with variable coefficients as
\begin{equation}
\label{gen}
\ddot{\phi}+f_1(t)\dot{\phi}+ f_2(t)\phi+f_3(t)\phi^n=0,
\end{equation}
where $f_i$ are functions of $t$ and $n \in {\cal Q}$ is a constant. Overhead dot represents differentiation with respect to cosmic time, $t$. Using Euler and Duarte’s \cite{duarte, euler, euler1} work on the integrability of the general anharmonic oscillator equation and the more applicable reproduction by Harko, Lobo and Mak\cite{harko}, this equation can be integrated under certain conditions. The essence of the integrability criterion is that, an equation of the form Eq.(\ref{gen}) can be point transformed into an integrable form iff $n\notin \left\{-3,-1,0,1\right\} $, provided the coefficients of Eq. (\ref{gen}) satisfy the differential condition

\begin{eqnarray}\nonumber\nonumber
\label{int-gen}
&& \frac{1}{n+3}\frac{1}{f_{3}(t)}\frac{d^{2}f_{3}}{dt^{2}} - \frac{n+4}{\left( n+3\right) ^{2}}\left[ \frac{1}{f_{3}(t)}\frac{df_{3}}{dt}\right] ^{2} \\&&
+ \frac{n-1}{\left( n+3\right) ^{2}}\left[ \frac{1}{f_{3}(t)}\frac{df_{3}}{dt}\right] f_{1}\left( t\right) \\&&
+ \frac{2}{n+3}\frac{df_{1}}{dt}+\frac{2\left( n+1\right) }{\left( n+3\right) ^{2}}f_{1}^{2}\left( t\right)=f_{2}(t).
\end{eqnarray}
Introducing a pair of new variables $\Phi$ and $T$ given by 
\begin{eqnarray}
\label{Phi}
\Phi\left( T\right) &=&C\phi\left( t\right) f_{3}^{\frac{1}{n+3}}\left( t\right)
e^{\frac{2}{n+3}\int^{t}f_{1}\left( x \right) dx },\\
\label{T}
T\left( \phi,t\right) &=&C^{\frac{1-n}{2}}\int^{t}f_{3}^{\frac{2}{n+3}}\left(
\xi \right) e^{\left( \frac{1-n}{n+3}\right) \int^{\xi }f_{1}\left( x
\right) dx }d\xi ,\nonumber\\
\end{eqnarray}%
where $C$ is a constant, Eq.(\ref{gen}) can then be written in an integrable form as 
\begin{equation}
\label{Phi1}
\frac{d^{2}\Phi}{dT^{2}}+\Phi^{n}\left( T\right) = 0.
\end{equation}

We focus our investigation on a particular case of polynomial coupling, i.e., where $\xi(\phi) = \xi_{0} \frac{\phi^2}{2}$. We also take the self-interaction potential $V(\phi) = V_{0} \frac{\phi^{(n+1)}}{(n+1)}$. Both positive powers and inverse powers of $\phi$ are very useful in the cosmological setting, in particular, inverse power law models are extremely useful as quintessence fields among other interesting properties. With this assumption, the scalar field evolution equation becomes
\begin{equation}\label{scalarKG1}
\ddot{\phi} +3H\dot{\phi} + 24 \xi_{0} \phi(H^4+H^2\dot{H}) + V_{0} \phi^n = 0.
\end{equation}
One can easily identify $f_{1}(t) = 3H$, $f_{2}(t) = 24 \xi_{0} (H^4 + H^2 \dot{H})$ and $f_{3} = V_{0}$.  \\

Provided $n\notin \left\{-3,-1,0,1\right\}$, the integrability criterion produces a differential equation for $H = \frac{\dot{a}}{a}$ given by
\begin{equation}
24 \xi_{0} H^4 - 18 \frac{(n+1)}{(n+3)^2} H^2 + \dot{H} (24 \xi_{0} H^2 - \frac{6}{(n+3)}) = 0,
\end{equation}
which we rewrite in the form
\begin{equation}
\dot{H} = \frac{18\frac{(n+1)}{(n+3)^2} - 24 \xi_{0} H^2}{24 \xi_{0} - \frac{6}{(n+3) H^2}}.
\end{equation}
During the final stages of the evolution of the collapsing scalar field, it is expected that the proper volume is very small and sharply decreasing in nature. Moreover, the rate of collapse must also be increasing rapidly. With that in mind one can say, $H = \frac{\dot{a}}{a} >> 0$ and is a sharply increasing function of time. Therefore, $\frac{1}{H^2}$ can be neglected compared to other term on the denominator on the RHS. With that simplification and the fact that $\dot{a} < 0$ for a collapsing model, we can write a solution for the scale factor, defining the time evolution of the collapsing scalar field as
\begin{eqnarray}\nonumber
\label{solution}
&& a(t) = \frac{1}{6(1+n)} e^{\frac{\sqrt{3(1+n)}}{2(3+n)\sqrt{\xi_{0}}}(t_{0}-t)} \\&&
- 2 \xi_{0} a_{0} (3+n)^2 e^{-\frac{\sqrt{3(1+n)}}{2(3+n)\sqrt{\xi_{0}}}(t_{0}-t)}.
\end{eqnarray}
Here both $t_{0}$ and $a_{0}$ are constants of integration and serves as the initial condition of the model alongwith the choice of self-interaction potential (value of $n$). In order to have a real time evolution one must enforce the restrictions $n > -1$ and $\xi_{0} > 0$.   \\

The time of reaching the zero proper volume can be calculated from equation (\ref{solution}) as
\begin{equation}\label{ts}
t_{s} = t_{0} - \frac{2(3+n)\sqrt{\xi_{0}}}{\sqrt{3(1+n)}} ln [2\sqrt{3 a_{0} \xi_{0} (1+n)} (3+n)].
\end{equation}

\section{Evolution of the Scale Factor for different initial conditions}

\subsection{Evolution of the scale factor for $a_{0} > 0$}
We present different possible outcomes of the collapse graphically and study the evolution varying different initial conditions.

\begin{figure}[h]
\begin{center}
\includegraphics[width=0.40\textwidth]{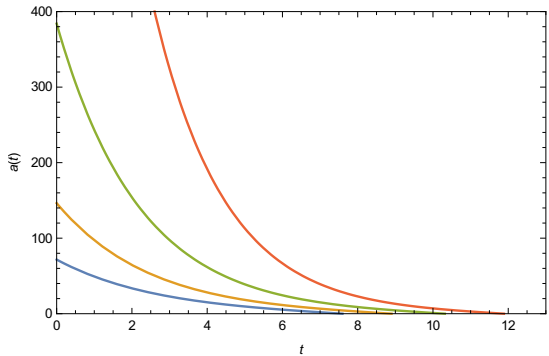}
\caption{Evolution of the scale factor with time for $V(\phi) = \frac{\phi^4}{4}$ and for different positive values of $a_{0}$. (Colour Code : $Blue \rightarrow a_{0} = 0.6$, $Yellow \rightarrow a_{0} = 0.5$, $Green \rightarrow a_{0} = 0.4$ and $Red \rightarrow a_{0} = 0.3$). For all values of the initial parameter $a_{0}$, the collapse reaches a zero proper volume, however the rate of collapse depends on the choice of the parameter.}
\end{center}
\label{fig:ltb1}
\end{figure}

In figure $1$, the scale factor is plotted as a function of time for a fixed value of $t_{0}$, for different positive values of $a_{0}$ and $n = 3$, i.e., $V(\phi) = \frac{\phi^4}{4}$. The evolution shows a rapid collapsing behavior For all values of $a_{0}$, and an ultimate zero proper volume end-state, however, the rapidity of the collapse and the time of formation of zero proper volume depends on the choice of the parameter. \\

An important note to make here is that, for the particular case of $n = 3$, i.e., $V(\phi) = \frac{\phi^4}{4}$, $t_{0} = 20$ and as long as a positive choice of $a_{0}$ is made, $\xi_{0}$ must be taken in between $0$ and $1$. For $\xi_{0} < 0$, there is no real evolution and for $\xi_{0} > 1$, the evolution becomes negative. Therefore a condition of $0 < \xi_{0} \leq 1$ must be enforced upon the strength of the coupling. However, this range can differ for a different set of initial conditions, i.e., for a different set of $n$ and $t_{0}$. We have presented a particular case with the note that for any such model, the strength of the coupling is very important and therefore the restriction over the allowed domain of $\xi_{0}$ must be accounted for.

\begin{figure}[h]
\begin{center}
\includegraphics[width=0.35\textwidth]{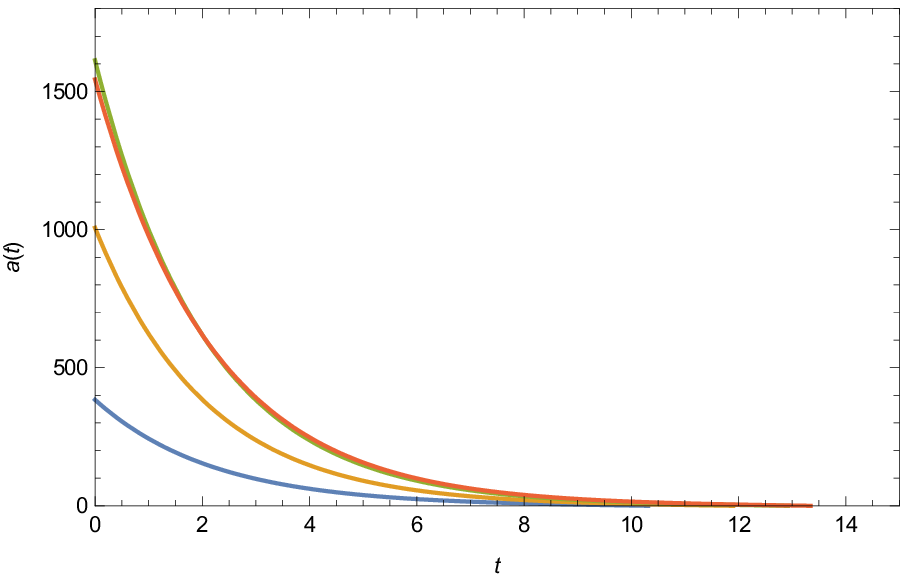}
\includegraphics[width=0.35\textwidth]{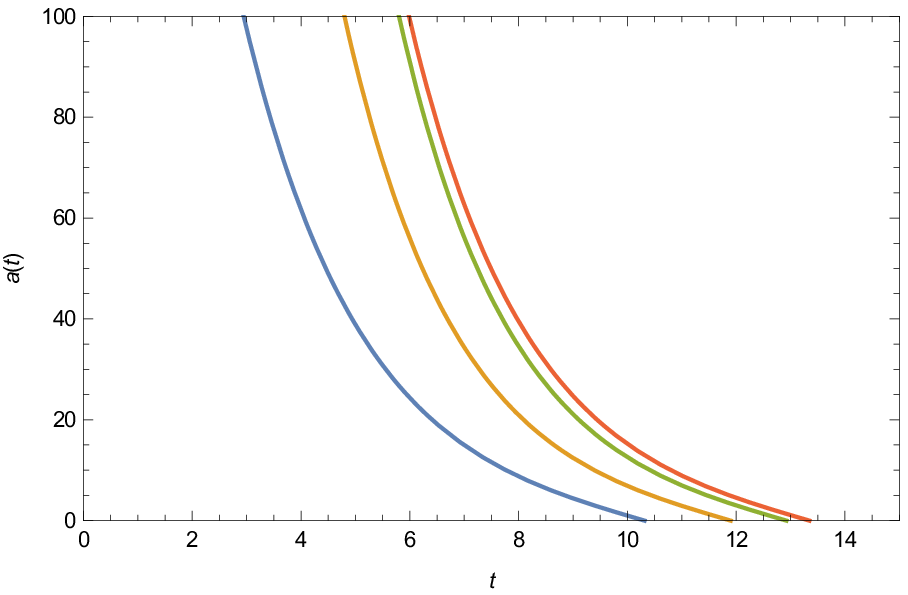}
\caption{Evolution of the scale factor with time for different self-interaction potential defined by $V(\phi) = \frac{\phi^{(n+1)}}{(n+1)}$ and for fixed $a_{0}$ and $\xi_{0}$. $(Colour Code : Blue \rightarrow n = 3$, $Yellow \rightarrow n = 1.5$, $Green \rightarrow n = 0.5$ and $Red \rightarrow n = 0.001)$. The two different plots are for different ranges.}
\end{center}
\label{fig:ltb1}
\end{figure}

In figure $2$, we plot the collapsing behavior for $\xi_{0} = 0.4$ and for different choices of $n$, i.e., for different choice of self interaction potential defined by $n = 3, 1.5, 0.5$ and $0.001$. For all the cases, the evolution starts at a finite value of the scale factor (as shown in the first figure) and a zero proper volme is reached at a finite future. The time of formation of singularity changes depending on the choice of $n$ (as shown by the figure below).

\begin{figure}[h]
\begin{center}
\includegraphics[width=0.35\textwidth]{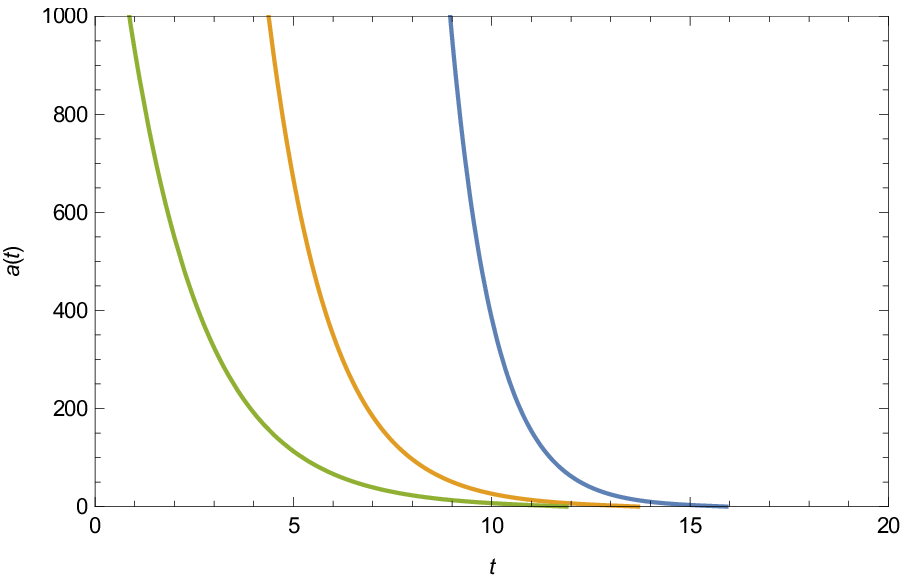}
\includegraphics[width=0.35\textwidth]{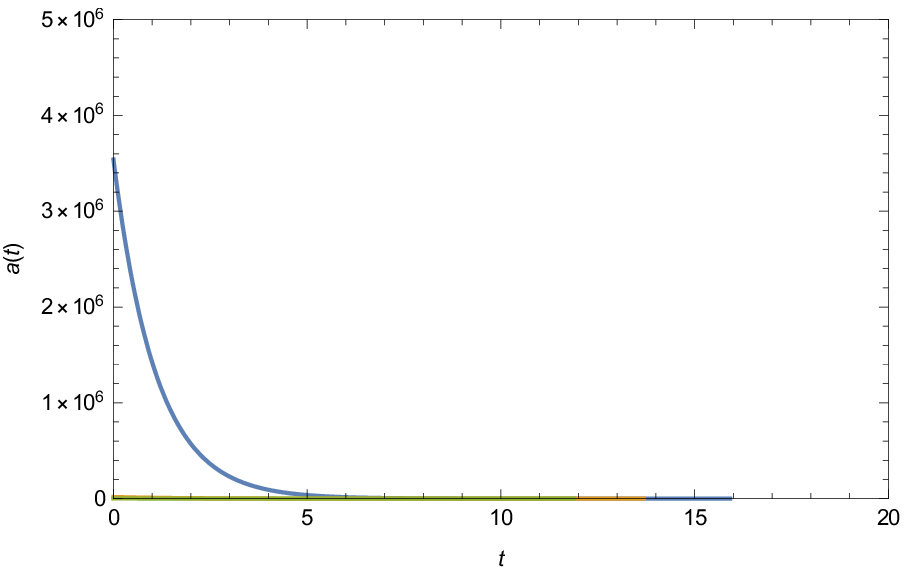}
\caption{Evolution of the scale factor with time for $V(\phi) = \frac{\phi^4}{4}$, fixed $a_{0}$ and different values of $\xi_{0}$.}
\end{center}
\label{fig:ltb1}
\end{figure}

In figure $3$, we show the evolution of the scale factor with time for different choice of the coupling $\xi_{0}$, fixing other initial parameters. For different value of $\xi_{0}$ the collapse ends up in a zero proper volume afterall, but for different $t_{s}$, which is also evident from equation (\ref{ts}). 

\subsection{Evolution of the scale factor for $a_{0} < 0$}

\begin{figure}[h]
\begin{center}
\includegraphics[width=0.35\textwidth]{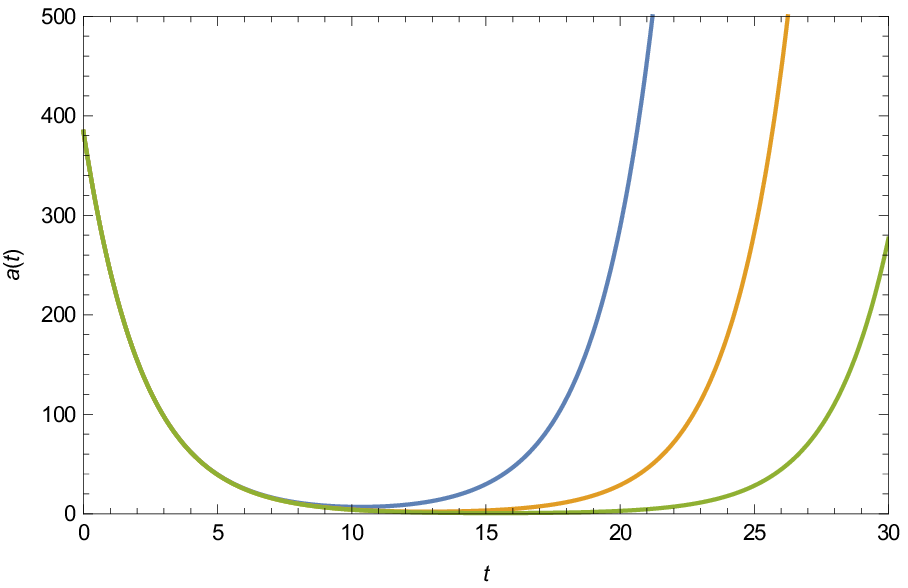}
\includegraphics[width=0.35\textwidth]{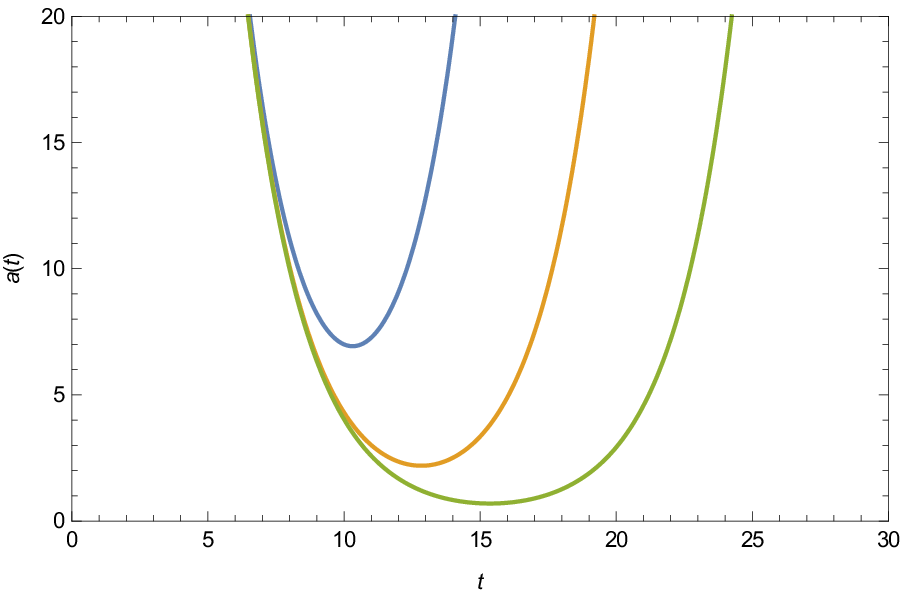}
\caption{Evolution of the scale factor with time for $V(\phi) = \frac{\phi^4}{4}$, $\xi_{0} = 0.4$ and for different negative values of $a_{0}$. $(Colour Code : Blue \rightarrow a_{0} = -10$, $Yellow \rightarrow a_{0} = -1$, $Green \rightarrow a_{0} = -0.1$ and $Red \rightarrow a_{0} = -0.00001)$. The evolution suggests that the scalar field experiences a collapse initially, only until a critical point after which the collapse changes into an expanding phase.}
\end{center}
\label{fig:ltb2}
\end{figure}

However, depending on the initial conditions, the collapse may not always lead to a zero proper volme. As shown in figure $4$, for all negative values of $a_{0}$, the system experiences a collapse initially, but only until a critical point (a non-zero minimum radius) after which it can not shrink further and experiences rapid expansion. The plot here is for $V(\phi) = \frac{\phi^4}{4}$ and $\xi_{0} = 0.4$. This behavior is valid for all negative values of $a_{0}$.

\begin{figure}[h]
\begin{center}
\includegraphics[width=0.35\textwidth]{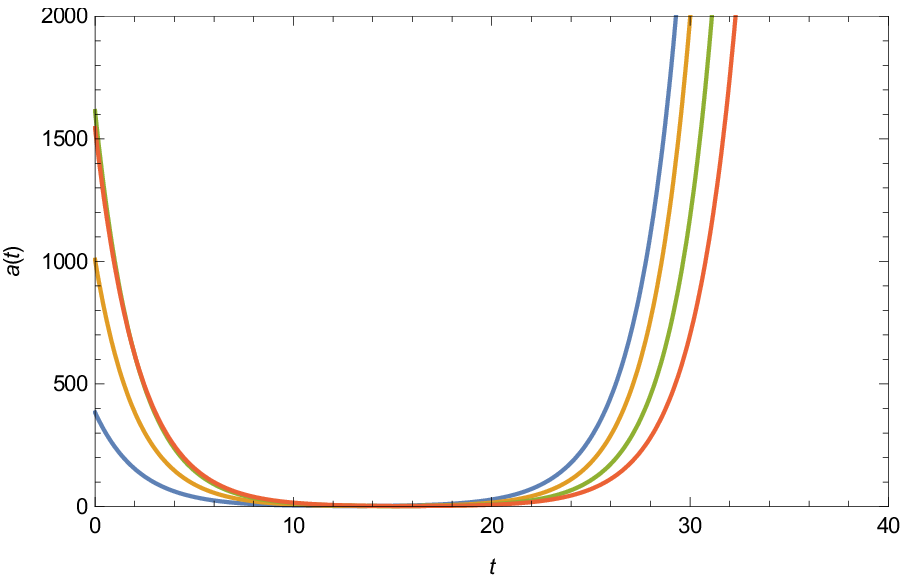}
\includegraphics[width=0.35\textwidth]{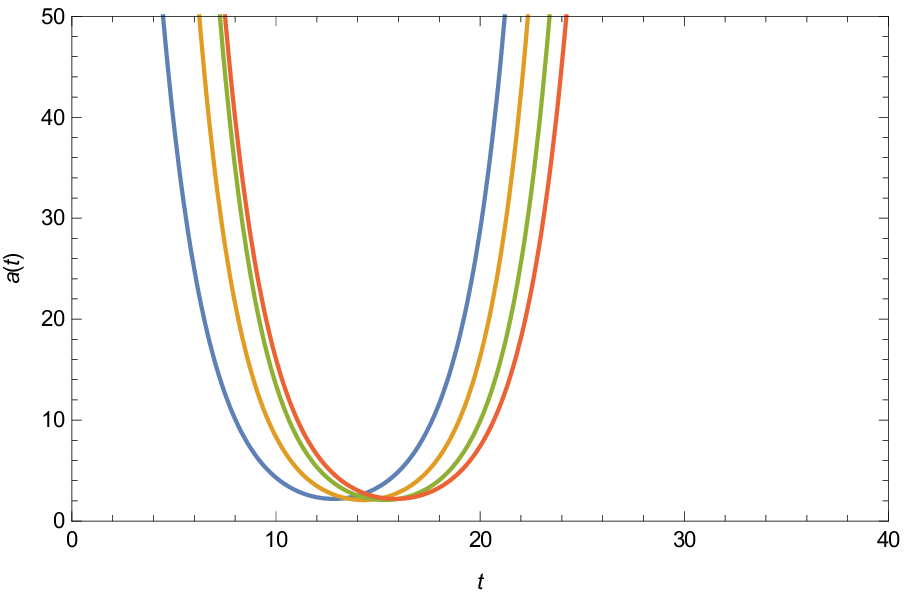}
\caption{Evolution of the scale factor with time for different $n$, i.e., for different choices of $V(\phi) = \frac{\phi^{(n+1)}}{n+1}$, $\xi_{0} = 0.4$ and for a particular negative value of $a_{0} = -1$. $(Colour Code : Blue \rightarrow n = 3$, $Yellow \rightarrow n = 1.5$, $Green \rightarrow n = 0.5$ and $Red \rightarrow n = 0.001)$.} 
\end{center}
\label{fig:ltb2}
\end{figure}

We present the evolution graphically for different choices of the self-interaction potential (depending on the choice of $n$) and for a particular choice of negative $a_{0}$, and fixed $\xi_{0}$. The evolution shown in figure $5$ suggests that the scalar field experiences a collapse initially, only until a critical point after which it experiences a bounce. The overall qualitative behavior remains the same for different $n$, however, there is an indication that, depending on $n$, the bouncing behaviour may be scaled.

\begin{figure}[h]
\begin{center}
\includegraphics[width=0.40\textwidth]{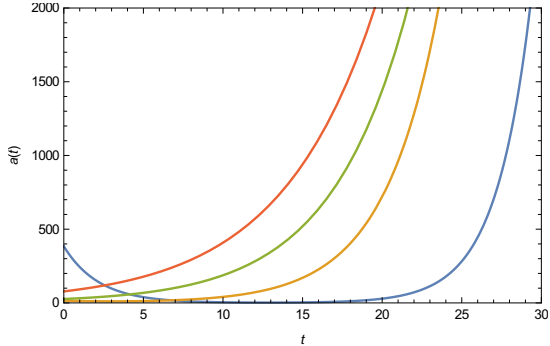}
\caption{Evolution of the scale factor with time for $V(\phi) = \frac{\phi^4}{4}$, $a_{0} = -1$ and for a different values of $\xi_{0}$. $(Colour Code : Blue \rightarrow \xi = 0.4$, $Yellow \rightarrow \xi = 1$, $Green \rightarrow \xi = 2$ and $Red \rightarrow \xi = 3)$.} 
\end{center}
\label{fig:ltb2}
\end{figure} 

In figure $6$, we plot the scale factor as a function of time for $V(\phi) = \frac{\phi^4}{4}$, $a_{0} = -1$ and for a different values of the coupling parameter $\xi_{0}$. The evolution suggests that the overall qualitative behavior may change depending on the value of $\xi_{0}$, in the sense that, the phase of an initial contraction of the scale factor depends on the choice of $\xi_{0}$. For instance, for $\xi_{0} = 0.4$, (shown by the blue curve) there is an initial collapsing phase before reaching an eventual non zero minimum cutoff, and the expanding phase begins thereafter. However, if one increases the value of $\xi_{0}$ gradually, it can be seen (Yellow for $\xi = 1$, Green for $\xi = 2$ and Red for $\xi = 3$) that the initial collapsing phase seems to become negligible and the entire solution turns into an expanding solution. 

\section{Evolution of the Scalar Field}

Using the defined point transformations (\ref{Phi}) and (\ref{T}), a general solution of the scalar field can be constructed from the transformed integrable form of the scalar field evolution equation (\ref{Phi1}). The solution is given in the form
\begin{equation}
\frac{dT}{d\Phi} = \frac{1}{\sqrt{2 \Big(C_{0} - \frac{\Phi^{(n+1)}}{(n+1)} \Big)}}.
\end{equation}

Calculating the coefficients $f_{1}(t) = 3 H$, $f_{2}(t) = 24 \xi_{0} (H^4 + H^2 \dot{H})$ and $f_{3} = V_{0}$ using the solution for scale factor (\ref{solution}), one appears at a differential equation governing the behaviour of the scalar field itself given by
\begin{equation}
\Big(\dot{\phi} + \frac{2\phi}{(n+3)} \Big)^2 = \frac{C^{-(n+1)} a^{-6\frac{(n+1)}{(n+3)}}}{2C_{0} + \frac{2}{(n+1)}\Big[C \phi a^{\frac{6}{(n+3)}}\Big]^{(n+1)}}.
\end{equation}

It is extremely non-trivial to solve this equation analytically without any choice of $n$. We present a particular case here. Using $C = 1$, $C_{0} = 0$ and investigating the equation numerically for $n = 3$ (i.e., for $V(\phi) = \frac{\phi^4}{4}$), we numerically solve the equation and the solution is written as
\begin{equation}\label{intescalar}
\phi(t) = e^{-\frac{1}{3}} \Bigg[C_{1} + 3 \int {\frac{g_{1}(t)}{(g_{2}(t) - \delta)^{4}}} \Bigg].
\end{equation}
Here, $g_{1}(t)$ and $g_{2}(t)$ are functions of time which we have written in this form for the sake of brevity. \\

ALthough it would have been better if a neat and closed for of the evolution could be written, however, the numerical integration of the equation (\ref{intescalar}) produces a very large expression for the scalar field (about $145$ terms). We present in brief the functional form so as to give an idea regarding how the scalar field can, in principle, evolve.

\begin{eqnarray} \nonumber
&& \phi(t) = \psi(t)^{\frac{1}{3}}, \\&& \nonumber
\psi(t) = s_{0} e^{-t} + \frac{g(t)}{j(t)}, \\&& \nonumber
j(t) = s_{1} \left(1728 b e^{\frac{\sqrt{\frac{1}{m}}t}{\sqrt{3}}} m-e^{\frac{d \sqrt{\frac{1}{m}}}{\sqrt{3}}}\right)^{3}, \\&& \nonumber \nonumber
g(t) \sim \Bigg[a_{1}e^{a_{2}-t}\Bigg(a_{i} + a_{j}e^{a_{k}+a_{l}t}\Bigg)\Bigg] + b_{i}e^{b_{j} + b_{k}t}\\&& \nonumber
{_2}F_1\left(1,2+\sqrt{3\xi_{0}};3+\sqrt{3\xi_{0}};1728 \xi_{0} a_{0} e^{\frac{1-t_{0}}{\sqrt{3\xi_{0}}}}\right) \\&& \nonumber
+ c_{i}e^{c_{j} + c_{k}t} {_2}F_1\left(1,2+\sqrt{3\xi_{0}};3+\sqrt{3\xi_{0}};1728 \xi_{0} a_{0} e^{\frac{t-t_{0}}{\sqrt{3\xi_{0}}}}\right).
\end{eqnarray}
All the parameteres $a_{1}$, $a_{2}$, $a_{i}-$s, $a_{j}-$s etc, are infact defined in terms of the parameters of the theory, i.e., $\xi_{0}$, $n$ and $a_{0}$. The function $g(t)$ consists of a total of $(80+33+31)$ terms where there are $80$ terms of the form of $\Bigg(a_{i} + a_{j}e^{a_{k}+a_{l}t}\Bigg)$, $33$ terms of the form of $b_{i}e^{b_{j} + b_{k}t}$ and $31$ terms of the form of $c_{i}e^{c_{j} + c_{k}t}$. It is quite obvious that a numerical examination is absolutely necessary over the time evolution of the scalar field. We note that, the parameters $\xi_{0}$ (i.e., the strength of the coupling of scalar field with the Gauss-Bonnet term) and the constant of integration $C_{1}$ play an important part in determining the behavior of the scalar field afterall. In the next subsection we present the numericla results of the evolution of the scalar field as a function of time for different parameters. 

\subsection{Evolution of the scalar field for $a_{0} > 0$, i.e., for collapsing scale factor}

\begin{figure}[h]
\begin{center}
\includegraphics[width=0.40\textwidth]{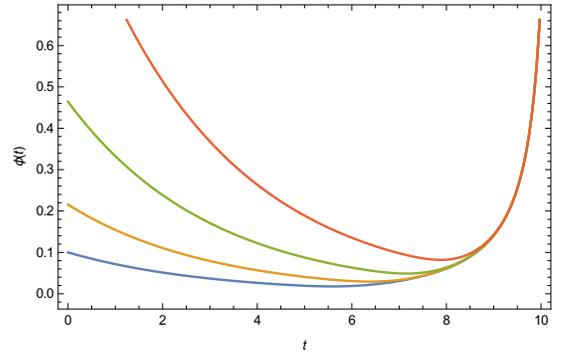}
\caption{Scalar field as a function of time for $V(\phi) = \frac{\phi^4}{4}$, $\xi_{0} = 0.4$ and for different values of $C_{1}$ ($Color Code : Blue \Rightarrow C_{1} = 0.001, Yellow \Rightarrow C_{1} = 0.01, Green \Rightarrow C_{1} = 0.1$ and $Red \Rightarrow 1.0$).}
\end{center}
\label{fig:ltb4}
\end{figure}

In figure $7$, the scalar field is plotted as a function of $t$ for $V(\phi) = \frac{\phi^4}{4}$, $\xi_{0} = 0.4$ and for different values of $C_{1}$ ($C_{1} = 0.001, C_{1} = 0.01, C_{1} = 0.1 and 1.0$). The scalar field diverges around the time of formation of singularity. However, depending on the value of $C_{1}$, the nature of the evolution before reaching singularity may be a little different but the qualitative behavior is this; the time evolution of the scalar field starts at a finite value and then decreases more-or-less steadily, before reaching a minimum critical value. Thereafter, the scalar field increases with time rapidly and diverges.

\begin{figure}[h]
\begin{center}
\includegraphics[width=0.40\textwidth]{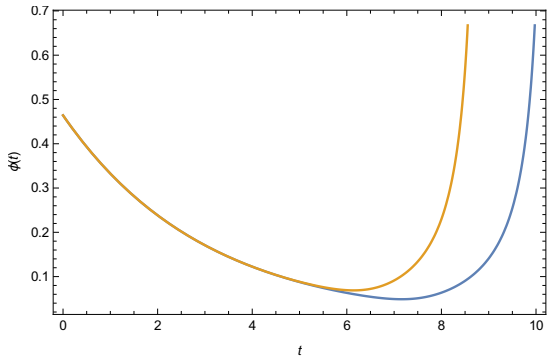}
\caption{Evolution of the scalar field as a function of time for $V(\phi) = \frac{\phi^4}{4}$, $C_{1} = 0.01$ and for different choices of coupling parameter $\xi_{0} = 0.4 (Blue)$ and $\xi_{0} = 0.5 (Yellow)$. }
\end{center}
\label{fig:ltb4}
\end{figure}

Evolution of the scalar field as a function of time is plotted in figure $8$ for $V(\phi) = \frac{\phi^4}{4}$, $C_{1} = 0.01$ and for different choices of $\xi_{0} = 0.4$ and $0.5$. For different values of the coupling parameter $\xi_{0}$, the scalar field diverges at different time. This is quite consistent because the time of formation of singularty depends on $\xi_{0}$. The same can be verified from equation (\ref{ts}) as well.

\subsection{Evolution of the scalar field for $a_{0} < 0$}
As discussed by figure $3$, $4$ and $5$, the evolution is not always collapsing forever, depending on the choice of the parameter $a_{0}$. For $a_{0} < 0$, the scale factor experiences a transition from a state of contraction into a rapid expansion. Here we show the behavior of the scalar field with time for such cases, i.e., for $a_{0} < 0$.  

\begin{figure}[h]
\begin{center}
\includegraphics[width=0.40\textwidth]{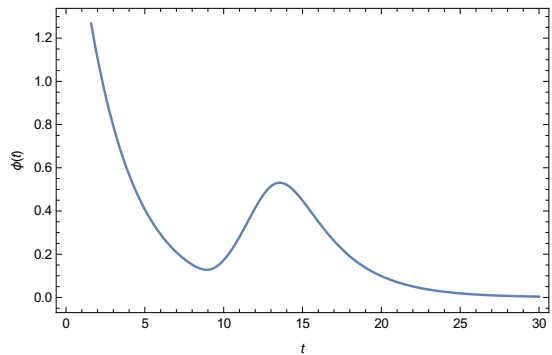}
\caption{Evolution of the scalar field as a function of time for $V(\phi) = \frac{\phi^4}{4}$, $C_{1} = 10$, $\xi_{0} = 0.4$ and $a_{0} = -1$.}
\end{center}
\label{fig:ltb4}
\end{figure}

In figure $9$, we have plotted $\phi(t)$ vs $t$ for $V(\phi) = \frac{\phi^4}{4}$, $C_{1} = 10$, $\xi_{0} = 0.4$ and $a_{0} = -1$. The scalar field starts at some finite value and gradually decreases, exhibiting some periodic behavior with time. Eventually it decays into a negligibly small positive value as shown in the figure.

\begin{figure}[h]
\begin{center}
\includegraphics[width=0.40\textwidth]{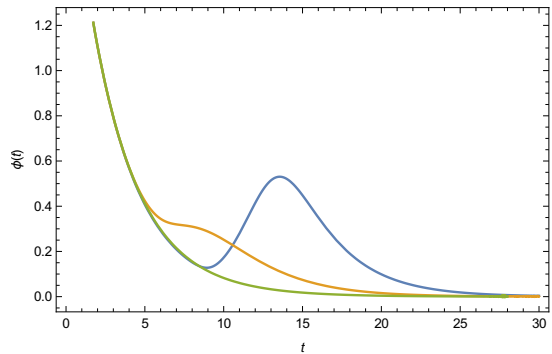}
\caption{Evolution of the scalar field as a function of time for $V(\phi) = \frac{\phi^4}{4}$, $C_{1} = 0.01$, $a_{0} = -1$ and for different choices of $\xi_{0} = 0.4 (Blue)$, $\xi_{0} = 1 (Yellow)$ and $\xi_{0} = 2 (Green)$.}
\end{center}
\label{fig:ltb4}
\end{figure}

However, the periodic time evolution of the scalar field is sensitive over the choice of $\xi_{0}$ as shown in figure $10$. The periodic nature seems to be absent as one gradually increases the value of $\xi_{0}$, (here, plotted for $\xi_{0} = 0.4, 1$ and $2$). 

\section{Evolution of the strong energy condition}
A collapsing perfect fluid is physically reasonable if it obeys the strong energy condition which is satisfied if for any timelike unit vector $w^{\alpha}$ and the following inequality holds
\begin{equation}
2 T_{\alpha\beta} w^{\alpha} w^{\beta} + T \geq 0,
\end{equation} 
where $T$ is the trace of the energy momentum tensor. The energy conditions were investigated in details for imperfect fluids by Kolassis, Santos and Tsoubelis \cite{kola}. Following their work, we investigate the validity of the strong energy condition ($(\rho+3p) > 0$) for our model. The strong energy condition can be violated only if the total energy density is negative or if there exists a large negative principal pressure of $T^{\alpha\beta}$.  \\

Since we have studied the solution for the scale factor from the scalar field evolution (\ref{scalarKG}) equation straightaway, the field equations (\ref{G00}) and (\ref{G11}) can be used to study the evolution of the constituent fluid density and pressure. We numerically study the strong energy condition and present it's nature by the following plots.  

\begin{figure}[h]
\begin{center}
\includegraphics[width=0.40\textwidth]{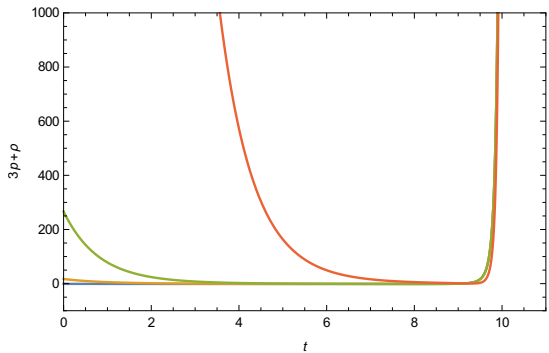}
\caption{Evolution of $(\rho+3p)$ with time for $V(\phi) = \frac{\phi^4}{4}$, $t_{0} = 20$, $a_{0} = 10$, $\xi_{0} = 0.4$ and for different choices of $C_{1}$ ($C_{1} = 0.1 (Blue)$, $C_{1} = 10 (Yellow)$, $C_{1} = 100 (Green)$ and $C_{1} = 10000 (Red)$).}
\end{center}
\label{fig:ltb4}
\end{figure}

In figure $11$, we plot the evolution of $(\rho+3p)$ as a function of time for a particular choice of potential $V(\phi) = \frac{\phi^4}{4}$, positive $a_{0}$ (ensuring the collapsing nature of the solution), $\xi_{0} = 0.4$ and for different choices of the initial conditon $C_{1}$ ($C_{1} = 0.1 (Blue), 10 (Yellow), 100 (Green)$ and $10000 (Red)$). The evolution suggests that that $(\rho+3p) \geq 0$ is satisfied throughout the evolution of the collapsing body, before reaching the curvature singularity where $(\rho+3p)$ increases sharply, as both pressure and density diverges eventually.

\begin{figure}[h]
\begin{center}
\includegraphics[width=0.40\textwidth]{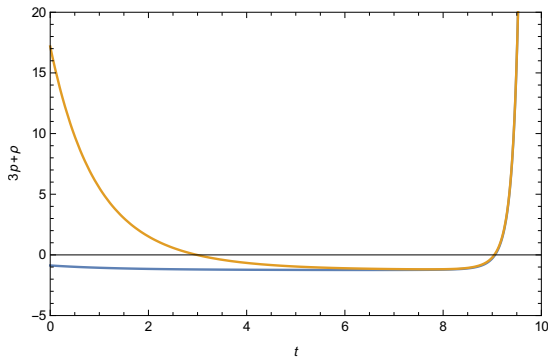}
\caption{Evolution of $(\rho+3p)$ with time for $V(\phi) = \frac{\phi^4}{4}$, $t_{0} = 20$, $a_{0} = 10$, $\xi_{0} = 0.4$ and for very small values of $C_{1}$ ($C_{1} = 0.001 (Blue)$ and $C_{1} = 0.0001 (Yellow)$.}
\end{center}
\label{fig:ltb4}
\end{figure}

However, depending on the value of $C_{1}$, we also find some cases where the strong energy condition may be violated during the collapsing evolution, before eventually diverging at the singular epoch. In figure $12$, we plot the evolution of $(\rho+3p)$ as a function time for $V(\phi) = \frac{\phi^4}{4}$, positive $a_{0}$, $\xi_{0} = 0.4$ and for very small positive values of $C_{1}$ ($C_{1} = 0.001 (Blue)$ and $C_{1} = 0.0001 (Yellow)$. \\

However, the evolution of the strong-energy condition in time leading to the conclusion that this is often violated, is not really an unexpected outcome in theories of gravity that contain strong-curvature terms. In the present case one can argue that the Gauss-Bonnett term in priciple can create an effective energy-momentum tensor whose contribution to the total energy-momentum tensor can lead to the violation of the strong-energy condition. Therefore, it may not be the nature of the perfect fluid afterall that violates the strong-energy condition. We also note here that the energy conditions of general relativity are a mathematical way of making the notion of locally positive energy density by stating that various linear combinations of the components of the energy momentum tensor must stay non-negative and it is sometimes argued that subtle quantum effects can violate all of the energy conditions. Moreover, there are examples of classical systems that violate all the energy conditions as well (For instance, Lorentzian-signature traversable wormholes \cite{barcelo}). The simplest possible source of classical energy condition violations is from the contribution of scalar fields, in particular, non-minimally coupled scalar field contributions, as worked out in details by Visser and Barcelo \cite{visser, barcelo}, Flanagan and Wald \cite{flanagan}.   \\

\begin{figure}[h]
\begin{center}
\includegraphics[width=0.40\textwidth]{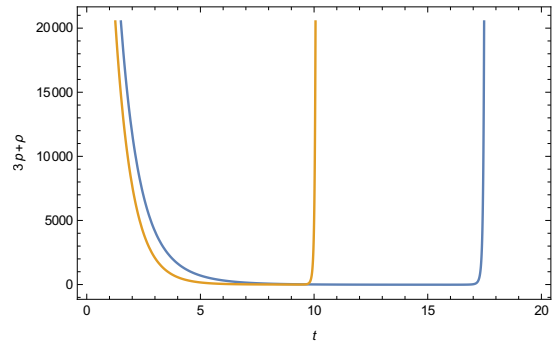}
\caption{\bf Evolution of $(\rho+3p)$ with time for $V(\phi) = \frac{\phi^4}{4}$, $t_{0} = 20$, $a_{0} = 10$ and for different value of $\xi_{0} = 0.04 (Blue)$ and $\xi_{0} = 0.4 (Yellow)$.}
\end{center}
\label{fig:ltb4}
\end{figure}

In figure $13$, the evolution of $(\rho+3p)$ with time is plotted for $V(\phi) = \frac{\phi^4}{4}$, positive $a_{0}$, $C_{1} = 10000$ and for different value of $\xi_{0} = 0.04 (Blue)$ and $\xi_{0} = 0.4 (Yellow)$. As can be seen from the graph, $(\rho+3p)$ maintains a positive signature throughout the collapse, however, for different value of $\xi_{0}$ the energy condition goes to positive infinity at different time. This is quite consistent, since we have already discussed that the strength of coupling with the Gauss-Bonnet term $\xi_{0}$ plays a crucial role in determining the time of formation of singularity.

\section{Nature and Visibility of the Singularity}

In order to investigate whether the singularity is a curvature singularity or just an artifact of coordinate choice, one must look into the behavior of the Kretschmann curvature ($K$) scalar at $t\rightarrow t_{s} = t_{0} - \frac{2(3+n)\sqrt{\xi_{0}}}{\sqrt{3(1+n)}} ln [2\sqrt{3 a_{0} \xi_{0} (1+n)} (3+n)]$, i.e., when the scale factor $a(t)$ goes to zero. For the metric presented in equation (\ref{metric}), $K$ has the expression,
\begin{equation}
 K = 6\bigg[\frac{\ddot{a}(t)^2}{a(t)^2} + \frac{\dot{a}(t)^4}{a(t)^4}\bigg]
 \label{curvature_scalar1}
\end{equation}

Using the solution of $a(t)$ (equation (\ref{solution})), it is straightforward to deduce that the Kretschmann scalar diverges at the zero proper volume and thus the collapsing body discussed here ends up in a curvature singularity.

\subsection{Formation of Apparent Horizon}

Whether the curvature singularity is visible to an exterior observer or not, depends on the formation of an apparent horizon. The condition for such a surface is given by
\begin{equation}
\label{app-hor}
g^{\mu\nu}R,_{\mu}R,_{\nu}=0,
\end{equation}
where $R$ is the proper radius of the two-sphere, given by $r a(t)$ in the present case.

Using the exact solution for collapse given by equations (\ref{solution}) and (\ref{app-hor}), we deduce the condition for formation of apparent horizon as

\begin{equation}
\frac{\sqrt{3}x}{12(3+n)\sqrt{(1+n)\xi_{0}}} + \frac{\sqrt{3 \xi_{0} (1+n)} a_{0} (3+n)}{x} - \lambda = 0,
\end{equation}

where $\lambda$ is a constant of separation and $x = e^{\frac{\sqrt{3(1+n)}}{2(3+n)\sqrt{\xi_{0}}}(t_{0}-t)}$. One can solve the above equation to exactly deduce the time of formtion of an apparent horizon as

\begin{eqnarray}
\label{tah}
&& t_{ap} = t_{0} \\&& \nonumber
- \frac{2(3+n) \sqrt{\xi_{0}}}{\sqrt{3(1+n)}} ln \Big[2(3+n) \sqrt{3(1+n)\xi_{0}} (\lambda \pm \sqrt{{\lambda}^2 - a_{0}})\Big].
\end{eqnarray} 

Comparing equation (\ref{tah}) with the time of formation of the singularity (\ref{ts}) we arrive at the very important expression
\begin{equation}\label{nakedcondition}
(t_{s} - t_{ap}) = \frac{2 (3+n) \sqrt{\xi_{0}}}{\sqrt{3(1+n)}} ln \Big[\frac{(\lambda \pm \sqrt{{\lambda}^2 - a_{0}})}{\sqrt{a_{0}}} \Big].
\end{equation}

To comment on the ultimate visibility of the collapse outcome, it is necessary to see if nonspacelike trajectories emanate from the singularity and reach a faraway observer. The singularity is at least locally naked if there are future directed nonspacelike curves that reach faraway observers. This is possible if the formation of apparent horizon is delayed or if there is no formation of horizon at all. In the present case the time of formation of singularity is independent of $r$ (given by equation (\ref{ts})) and therefore it is only natural that the entire collapsing system (scalar field and perfect fluid) would reach the singularity simulteneously at $t = t_{s}$. This kind of singularity is always expected to be covered by the formation of an apparent horizon as discussed by Joshi, Goswami and Dadhich \cite{naresh}. Here, from equation (\ref{nakedcondition}), it can be said that the formation of an ultimate covered singularity is dependent over initial conditions such as $\xi_{0}$, $\lambda$ and $a_{0}$ so that $(t_{s} - t_{ap}) > 0$.  
However, in principle, one can also ask about the possible end state if the initial conditions conspire to make the situation to end up otherwise, i.e., $(t_{s} - t_{ap}) < 0$. In such a case, there is no formation of apparent horizon at all, since all the collapsing shells labelled by different values of $r$, shrinks to zero proper volume and the physical quantities diverge at $t = t_{s}$, which is reached before $t_{ap}$. Therefore the singularity remains naked and the condition for such an end-state can be written from equation (\ref{nakedcondition}) as
\begin{equation}\label{nakedcondition2}
a_{0} = \frac{4 \lambda^2 \delta^2}{(1 + \delta)^2},
\end{equation} 
where $0 < \delta < 1$.

\section{Matching with an exterior Vaidya Spacetime}
For the sake of completeness, proper junction conditions are to be examined carefully which allow a smooth matching of an exterior geometry with the collapsing interior. First of all it was extensively shown by Goncalves and Moss \cite{gonca} that any sufficiently massive collapsing scalar field can be formally treated as collapsing inhomogeneous dust in general relativity. Moreover, astrophysical objects undergoing a gravitational collapse can be expected to be in an almost vacuum spacetime, and therefore the exterior spacetime around a spherically symmetric dying star is well described by the Schwarzschild geometry. From the continuity of the first and second differential forms, the matching of the sphere to a Schwarzschild spacetime on the boundary surface, $\Sigma$, is extensively worked out in literature \cite{santos, chan, kola2, maharaj}.         \\

However, conceptually this leads to an inconsistency, since Schwarzschild has zero scalar field. Therefore, such a matching would lead to a discontinuity in the scalar field, and a delta function in the gradient of
the scalar field. As a consequence, there will appear square of a delta function in the stress-energy, which is definitely an inconsistency. Since we have a scalar field distribution inside the collapsing sphere, it is more physical to match the interior with a Vaidya exterior solution (for a detailed analysis of vaidya matching in a scalar field collapse we refer to \cite{pankajritu1, pankajritu2}) across a boundary hypersurface defined by $\Sigma$. The metric just inside $\Sigma$ is,
\begin{equation}\label{interior}
d{s_-}^2=dt^2-a(t)^2dr^2-r^2 a(t)^2d{\Omega}^2,
\end{equation}
and the metric in the exterior of $\Sigma$ is given by
\begin{equation}\label{exterior}
d{s_+}^2=(1-\frac{2M(r_v,v)}{r_v})dv^2+2dvdr_v-{r_v}^2d{\Omega}^2.
\end{equation}

Matching the first fundamental form on the hypersurface we get
\begin{equation}\label{cond1}
\Big(\frac{dv}{dt} \Big)_{\Sigma}=\frac{1}{\sqrt{1-\frac{2M(r_v,v)}{r_v}+\frac{2dr_v}{dv}}}
\end{equation}
and
\begin{eqnarray}\nonumber
\label{cond2}
&& (r_v)_{\Sigma} = r a(t) \\&& \nonumber
= r \Bigg[\frac{1}{6(1+n)} e^{\frac{\sqrt{3(1+n)}}{2(3+n)\sqrt{\xi_{0}}}(t_{0}-t)} \\&& \nonumber
- 2a_{0}(3+n)^2 e^{-\frac{\sqrt{3(1+n)}}{2(3+n)\sqrt{\xi_{0}}}(t_{0}-t)}\Bigg].
\end{eqnarray}

Matching the second fundamental form yields,
\begin{equation}\label{cond3}
\Big(r a(t) \Big)_{\Sigma} = r_v\left(\frac{1-\frac{2M(r_v,v)}{r_v}+\frac{dr_v}{dv}}{\sqrt{1-\frac{2M(r_v,v)}{r_v}+\frac{2dr_v}{dv}}}\right)
\end{equation}
Using equations (\ref{cond1}), (\ref{cond2}) and (\ref{cond3}) one can write
\begin{equation}\label{dvdt2}
\left(\frac{dv}{dt} \right)_{\Sigma} = \frac{3r a(t)^2 - r^2}{3(ra(t)^2 - 2Ma(t))},
\end{equation}

where $a(t) = \frac{1}{6(1+n)} e^{\frac{\sqrt{3(1+n)}}{2(3+n)\sqrt{\xi_{0}}}(t_{0}-t)} - 2a_{0}(3+n)^2 e^{-\frac{\sqrt{3(1+n)}}{2(3+n)\sqrt{\xi_{0}}}(t_{0}-t)}$.

From equation (\ref{cond3}) one obtains
\begin{equation}\label{M}
M_{\Sigma} = \frac{1}{4} \Bigg[ra(t) + \frac{r^3}{9 a(t)^3} + \sqrt{\frac{1}{r a(t)} + \frac{r^3}{81 a(t)^9} - \frac{2r}{9 a(t)^5}} \Bigg].
\end{equation}
Matching the second fundamental form we can also write the derivative of $M(v,r_v)$ as
\begin{equation}\label{dM}
M{(r_v,v)}_{,r_v}=\frac{M}{r a(t)} - \frac{2r^2}{9 a(t)^{4}}.
\end{equation}
Equations (\ref{cond2}), (\ref{dvdt2}), (\ref{M}) and (\ref{dM}) completely specify the matching conditions at the boundary of the collapsing scalar field.

\section{Conclusion}
The present work has been entirely dedicated towards a deeper understanding of a self-interacting scalar field collapse in Scalar-Einstein-Gauss-Bonnet gravity. The scalar field couples non-minimally with the Gauss-Bonnet term by a term quadratic in the scalar field ($\xi_{0} \frac{\phi^2}{2}$). The possibility of the collapse reaching a zero proper volume singularity is seen to be dependent on the coupling parameter $\xi_{0}$, choice of self-interaction potential and most importantly the initial condition $a_{0}$ which can be related to the initial radius/volume of the collapsing star. We also comment on the allowed domain of the choice of the coupling parameter $\xi_{0}$ to have a real evolution of the collapse. It is observed that the collapse ends up in a curvature singularity, where the Kretschmann curvature scalar blows up, alongwith density and pressure of the constituent fluid and the scalar field. The strong energy condition is showed to be valid throughout the collapse. However, depending on certain initial conditions defining the initial distribution of the scalar field, the energy conditions may be violated, which can be attributed to the introduction of a non-minimal coupling of a scalar field in the lagrangian, which is known to be a possible classical system that can violate almost all of the energy conditions \cite{visser, barcelo}.  \\

For the sake of completeness we match the interior collapsing solution with the an exterior Vaidya
geometry on a boundary hypersurface, since the presence of a Gauss-Bonnet term non-minimally coupled to a scalar field generates a non-zero effective energy momentum tensor arising from spacetime curvature. Therefore matching with an exterior vacuum solution in the presence of Gauss-Bonnet term may lead to inconsistency. Very recently this has been investigated by Banerjee and Paul \cite{nbtp}. \\

The time of formation of the singularity is independent of $r$, which suggests that all the collapsing shells labelled by different values of $r$ collapses simulteneously when the zero proper volume is reached. Such a singularity is always expected to be covered by a horizon as far as similar studies under the domain of GR are concerned. We note here that in the present case of Scalar-Einstein-Gauss-Bonnet gravity with a polynomial coupling, the ultimate end-state of a covered singularity is conditionally consistent with the corresponding results in GR. For certain initial conditions (defined by equations (\ref{nakedcondition}) and (\ref{nakedcondition2})) there is a probability of an end state where there is no possibility of formation of horizon at all. \\

We also observe some interesting results, for instance, depending on the signature of the aforementioned parameter $a_{0}$, the evolution of scale factor suggests that all of the possible collapsing scenario may not lead to a curvature singularity. Rather, for $a_{0} < 0$, the fluid undergoes contraction only unto a minimum non-zero cut-off radius, after which it goes into a rapidly expanding phase. It is also seen that the scalar field itself in such cases, decreases monotonically with time, exhibits certain periodic behavior, before becoming negligibly small. This can be somewhat compared with the phenomena of collapse and dispersal for a scalar field in general relativity which have drawn considerable interest in recent years, mainly from a numerical perspective, subtly pointing towards the existence of a critical phenomena is the whole collapsing picture (For details, we refer to the monograph by Gundlach \cite{gund}). Recently Bhattacharya, Goswami and Joshi worked out a sufficient condition for the dispersal to take place for a collapsing scalar field in GR that initially begins with a contraction and showed that the transition of the collapsing body into expanding nature is crucially connected wiith the change of the gradient of the scalar field \cite{bgj}. In the present case we can note that the signature of the parameter $a_{0} > 0$ or $a_{0} < 0$ is the defining factor of the transition; i.e., whether or not the collapse will lead into a zero proper volume or a dispersal of scalar field shall take place after the scale factor reaches a minimum cut-off volume. Since $a_{0}$ comes from the expression defining the scale factor ($a(t) = \frac{1}{6(1+n)} e^{\frac{\sqrt{3(1+n)}}{2(3+n)\sqrt{\xi_{0}}}(t_{0}-t)} - 2a_{0}(3+n)^2 e^{-\frac{\sqrt{3(1+n)}}{2(3+n)\sqrt{\xi_{0}}}(t_{0}-t)}$), the initial volume of the collapsing system can be predicted as a defining factor connected to the value of $a_{0}$.  \\

We conclude with the note that this is indeed a simple case, in the sense that we have considered spatial homogeneity in the metric components as well as the scalar field. However, the solution found is simple enough to encourage further allied investigations in this direction, such as, the possibility of a collapse even when the energy conditions are violated may subtly direct one's attention towards a possible clustering of dark energy distribution. Apart from that, this work also helps further expansion of the usefulness and scope of the particular method of integrability of anharmonic oscillator used here. The theorem which inspires this method is self sufficient as was discussed by Euler \cite{euler1}. The same has been proved in the context of a massive scalar field minimally coupled to gravity by Chakrabari and Banerjee \cite{scnb2}, where the solutions found by virtue of this theorem indeed solve the Klein Gordon type evolution equation once they are put back in the equation. In the current context, the solutions are far more complicated to say the least, as is evident from the expression of the scalar field, as worked out in details in section $V$. Since the criterion for integrability was investigated under the expectation that the proper volume would be very small and sharply decreasing in nature so that one can neglect the $\frac{1}{H^2}$ term, to justify such an arguement one can put the solutions (\ref{solution}) and (\ref{intescalar}) together into the equation (\ref{scalarKG1}) and study for different initial conditions defining the scalar field. It can be confirmed that for relevant cases, the solutions put into the Klein-Gordon type equation (\ref{scalarKG1}) yields a number very close to zero ($\sim 10^{-8}$ or lower). This approach of point transforming the klein-gordon type equation for the scalar field, to extract the solution out of a seemingly impossible non-linear system has worked really well in the past while investigating different setups of massive scalar field collapse and also inspired a handy reconstruction technique which helps one to assess the particular form of the lagrangian of a modified theory of gravity \cite{scjs}. We note here that, although the assumption of integrability of the scalar field evolution equation is inspired only from a mathematical perspective, solutions found by means of this assumption are by no means unphysical. Used properly, this method can potentially be useful for allied investigations as well, for instance, a detailed definition of the possible bounds over the choice of coupling function $\xi$ (somewhat similar to a possible Higgs-Kreschmann invariant coupling in white dwarfs and neutron stars, as shown by Wegner and Onofrio \cite{ono1, ono2}). Further investigation under the setup of a Scalar-Einstein-Gauss-Bonnet gravity in a more generalized scenario (inclusion of factors spatial inhomogeneity, pressure anisotropy, heat flux etc) can be done using this method properly and more rigorously and will be reported elsewhere in future.  

\section{Acknowledgements}
The author would like to thank Professor Sayan Kar and Professor Narayan Banerjee for useful comments and suggestions. The author was supported by the National Post-Doctoral Fellowship (file number: PDF/2017/000750) from the Science and Engineering Research Board (SERB), Government of India.


\begin{thebibliography}{90}
 \bibitem{os}
  J. R. Oppenheimer, H. Snyder, Phys. Rev. {\bf 56}, 455 (1939).
 
 \bibitem{dutt}
  B. Datt, Z. Phys. {\bf 108}, 314 (1938); Reprinted as a Golden Oldie, Gen. Relativ.  Gravit., {\bf 31},  1615 (1999)
 
 \bibitem{rp}
  R. Penrose, Nuovo Cimento Rivista Serie, {\bf 1} (1969)

 \bibitem{thesis}
  S. Chakrabarti, arXiv:1709.01512v1 [gr-qc] 
 
 \bibitem{joshi1}
  P.S. Joshi, Global Aspects in Gravitation and Cosmology (Clarendon Press, Oxford, 1993).
 
 \bibitem{joshi2}
  P.S. Joshi. arXiv:1305.1005.
 
 \bibitem{M}
  N. Arkani-Hamed, S. Dimopoulos and G. Dvali, Phys. Lett. B. {\bf 429}, 263 (1998); I. Antoniadis, N. Arkani-Hamed, S. Dimopoulos and G. Dvali, Phys. Lett. B. {\bf 436}, 257 (1998); L. Randall and R. Sundrum, Phys. Rev. Lett. {\bf 83}, 3370 (1999); Phys. Rev. Lett. {\bf 83}, 4690 (1999); G. Dvali, G. Gabadadze, and M. Porrati, Phys. Lett. B {\bf 485}, 208 (2000); G. Dvali and G. Gabadadze, Phys. Rev. D. {\bf 63}, 065007 (2001); G. Dvali, G. Gabadadze, and M. Shifman, Phys. Rev. D. {\bf 67}, 044020 (2003); P. Hořava and E. Witten, Nucl. Phys. B. {\bf 475}, 94 (1996); A. Lukas, B. A. Ovrut, K.S. Stelle and D. Waldram, Phys. Rev. D. {\bf 59}, 086001 (1999); A. Lukas, B. A. Ovrut and D. Waldram, Phys. Rev. D. {\bf 60}, 086001 (1999).


 \bibitem{bdc}
  A. Banerjee, U. Debnath, and S. Chakraborty, Int. J. Mod. Phys. D. {\bf 12}, 1255 (2003).
 
 \bibitem{patil}  
  K. D. Patil, Phys. Rev. D. {\bf 67}, 024017 (2003).
  
 \bibitem{gosjos}  
  R. Goswami and P.S. Joshi Phys. Rev. D. {\bf 69}, 044002 (2004); Phys. Rev. D. {\bf 69}, 104002 (2004).
 
 \bibitem{superstring}
  D. J. Gross and J. H. Sloan, Nucl. Phys. B. {\bf 291}, 41 (1987); M. C. Bento and O. Bertolami, Phys. Lett. B. {\bf 368}, 198 (1996).
  
  \bibitem{nojiri1}
  S. Nojiri and S. D. Odintsov, Phys. Lett. B. {\bf 631}, 1 (2005).
  
  \bibitem{nojiri2}
  S. Nojiri, S. D. Odintsov and M. Sasaki, Phys. Rev. D. {\bf 71}, 123509 (2004).
  
  \bibitem{cognola}  
  G. Cognola, E. Elizalde, S. Nojiri, S. D. Odintsov and S. Zerbini, Phys. Rev. D. {\bf 73}, 084007 (2006).

  \bibitem{koiv1}   
  T. Kolvisto, D. Mota, Phys. Lett. B. {\bf 644}, 104 (2007).

  \bibitem{koiv2}   
  T. Kolvisto, D. Mota, Phys. Rev. D. {\bf 75}, 023518 (2007).

  \bibitem{boul1}
  D. G. Boulware and S. Deser, Phys. Rev. Lett. {\bf 55}, 2656 (1985).

  \bibitem{boul2}
  D. G. Boulware and S. Deser, Phys. Lett. B. {\bf 175}, 409 (1986).
  
  \bibitem{gurses}
  M. Gurses, Gen. Relativ. Gravit. {\bf 40} : 1825, (2008).
  
  \bibitem{zwi}
  B. Zwiebach, Phys. Lett. B. {\bf 156}, 315 (1985).
  
  \bibitem{zu}
  B. Zumino, Phys. Rep. {\bf 137}, 109 (1986).
  
  \bibitem{kawa1}
  S. Kawai, M. Sakagami and J. Soda, Phys. Lett. B. {\bf 437}, 284 (1998).
  
  \bibitem{soda}
  J. Soda, M. Sakagami and S. Kawai, arXiv:gr-qc/9807056.

  \bibitem{kawa2}  
  S. Kawai and J. Soda, Phys. Lett. B 460, 41 (1999).
  
  \bibitem{guo}
  Z. K. Guo and D. J. Schwarz, Phys. Rev. D. {\bf 81}, 123520 (2010).

  \bibitem{koh}
  S. Koh, B. H. Lee, W. Lee and G. Tumurtushaa, Phys. Rev. D. {\bf 90}, 063527 (2014).
  
  \bibitem{maeda1}
  H. Maeda, Class. Quantum Grav. {\bf 23}, 2155 (2006).

  \bibitem{maeda2}
  H. Maeda, Phys. Rev. D. {\bf 73}, 104004, (2006).
  
  \bibitem{mann1}
  T. Taves, C. D. Leonard, G. Kunstatter and R. B. Mann, Class. Quant. Grav. {\bf 29}, 015012 (2012).     
    
  \bibitem{gonca}
  S. M. C. V. Goncalves and I. G. Moss, Class. Quant. Grav. {\bf 14}, 2607 (1997).
 
  \bibitem{chop}
  M. W. Choptuik, Phys. Rev. Lett. {\bf 70}, 9 (1993).
  
  \bibitem{brady}
  P. R. Brady, Class. Quant. Grav. {\bf 11}, 1255 (1995).

  \bibitem{gund}
  C. Gundlach, Phys. Rev. Lett. {\bf 75}, 3214 (1995); C. Gundlach, Critical phenomena in Gravitational Collapse: Liv. Rev. Rel. {\bf 2} :4(1999).
  
  \bibitem{scnb1}
  N.Banerjee and S. Chakrabarti, Phys. Rev. D, {\bf 95}, 024015 (2017).
  
  \bibitem{mann2}
  N. Deppe, C. D. Leonard, T. Taves, G. Kunstatter and R. B. Mann, Phys. Rev. D. {\bf 86}, 104011, (2012).
    
  \bibitem{golod}
  S. Golod and T. Piran Phys. Rev. D. {\bf 85}, 104015 (2012).
  
  \bibitem{scalarcollapse}
  R. Goswami and P. S. Joshi, Phys. Rev. D. {\bf 65}, 027502 (2004); R. Giambo, Class. Quant. Grav. {\bf 22}, 2295 (2005); S. Goncalves, Phys. Rev. D. {\bf 62}, 124006 (2000); R. Goswami and P. S. Joshi, Mod. Phys. Lett. A. {\bf 22}, 65 (2007). 
  
  \bibitem{nbtp}
  N. Banerjee and T. Paul, arXiv:1709.07271v1 [gr-qc].
  
  \bibitem{doneva1}
  D. Doneva and S. Yazadjiev, arXiv:1711.01187v1 [gr-qc].
  
  \bibitem{doneva2}
  D. Doneva and S. Yazadjiev,  arXiv:1712.03715v1 [gr-qc].
  
  \bibitem{silva}
  H. O. Silva, J. Sakstein, L. Gualtieri, T. P. Sotiriou and E. Berti, arXiv:1711.02080v2 [gr-qc]. 
  
  \bibitem{kanti0}  
  G. Antoniou, A. Bakopoulos and P. Kanti, arXiv:1711.03390v2 [hep-th].

  \bibitem{kanti00}  
  G. Antoniou, A. Bakopoulos and P. Kanti, arXiv:1711.07431v1 [hep-th].

  \bibitem{kanti}
  P. Kanti, R. Gannouji and N. Dadhich, Phys. Rev. D. {\bf 92}, 041302 (2015).

    
  \bibitem{scnb2}
  S. Chakrabarti and N. Banerjee, Eur. Phys. J. C. {\bf 77} : 166 (2017).
  
  \bibitem{duarte}
  L. G. S. Duarte, I. C. Moreira, N. Euler and W. H. Steeb, Physica Scripta. {\bf 43}, 449, (1991).
  
  \bibitem{euler}
  N. Euler, W. H. Steeb and K. Cyrus, J. Phys. A. Math. Gen. {\bf 22}, L195 (1989).
    
  \bibitem{scjs}
  S. Chakrabarti, J. L. Said, G. Farrugia, Eur. Phys. J. C. {\bf 77} : 815 (2017).
  
  \bibitem{euler1}
  N. Euler, Journal of Nonlinear Mathematical Physics, {\bf 4}, 310 (1997).

  \bibitem{harko}
  T. Harko, F. S. N. Lobo and M. K. Mak, Journal of Pure and Applied Mathematics: Advances and Applications {\bf 10} (1) 115 (2013).
  
  \bibitem{kola}
  C. A. Kolassis, N. O. Santos and D. Tsoubelis, Class. Quant. Grav. {\bf 5}, 1329 (1988).  

  \bibitem{visser}
  M. Visser and C. Barcelo, COSMO-99: pp. 98 (2000).
  
  \bibitem{flanagan}
  E. Flanagan and R. Wald, Phys. Rev. D. {\bf 54} : 6233, (1996).
    
  \bibitem{barcelo}
  C. Barcelo and M. Visser, Phys. Lett. B. {\bf 466}, 127 (1999).
  
  \bibitem{naresh}
  P. S. Joshi, R. Goswami and N. Dadhich, Phys. Rev. D, {\bf 70}, 087502 (2004).
  
  \bibitem{santos}
  N. O. Santos, Mon. Not. R. Astron. Soc. {\bf 216}, 403 (1985).
  
  \bibitem{chan}
  R. Chan, Mon. Not. R. Astron. Soc. {\bf 316}, 588 (2000).
  
  \bibitem{kola2}
  C. A. Kolassis, N. O. Santos, D. Tsoubelis. Astrophysical Journal, Part 1 (ISSN 0004-637X), {\bf 327},
755, (1988).
  
  \bibitem{maharaj}
  S. D. Maharaj, M. Govender, Int. J. Mod. Phys. D. {\bf 14}, 667 (2005).
  
  \bibitem{pankajritu1} 
  P. S. Joshi, R. Goswami, Phys. Rev. D. {\bf 69} 064027 (2004).

  \bibitem{pankajritu2} R. Goswami and P. S. Joshi, Phys. Rev. D; {\bf 65}, 027502 (2004).  
  
  \bibitem{bgj}
  S. Bhattacharya, R. Goswami and P. S. Joshi, Int. J. Mod. Phys. D. {\bf 20}, No. 6, 1123 (2011). 

  \bibitem{ono1}
  R. Onofrio and G. A. Wegner, Astrophys. J. {\bf 791} : 125, (2014).
 
  \bibitem{ono2}
  G. A. Wegner and R. Onofrio, Eur. Phys. J. C. {\bf 75} : 307, (2015).   
  
  
\end{thebibliography}
\end{document}